\documentclass[12pt]{article}

\usepackage{graphicx}
\usepackage{comment}
\usepackage{color} 
\usepackage{slashed}
\usepackage{amsmath}
\usepackage[pdfstartview=FitH,colorlinks=true,linkcolor=black,anchorcolor=black,citecolor=black,urlcolor=black]{hyperref}
\usepackage{amsmath,amssymb,titling,authblk}
\usepackage{slashed}
\usepackage{amsmath}
\usepackage{amscd}
\usepackage[normalem]{ulem}
\usepackage{appendix}
\usepackage{bbold}

\usepackage[utf8]{inputenc}
\usepackage{slashed}
\usepackage[top=2.3cm,right=2.3cm,left=2.3cm,bottom=2.3cm]{geometry}

\definecolor{verde}{cmyk}{.83,.21,1,.08}
\definecolor{darkorchid}{rgb}{0.6, 0.2, 0.8}
\definecolor{darkgreen}{rgb}{0,.5,0}

\def\({\left(}
\def\){\right)}
\def\[{\left[}
\def\]{\right]}

\newcommand{\be}{\begin{equation}}
	\newcommand{\ee}{\end{equation}}
\newcommand{\bea}{\begin{eqnarray}}
	\newcommand{\eea}{\end{eqnarray}}
\newcommand{\e}{\mathrm{e}}

\begin{document}
	
	\title{\Large\bf Poisson electrodynamics on $\kappa$-Minkowski space-time}
	
	\author[1]{O. Abla}
	\author[2]{M. J. Neves}
	\affil[ ]{}
	\affil[1]{\textit{\footnotesize CCNH - Universidade Federal do ABC, 09210-580, Santo Andr\'e, SP, 
			Brazil. }}
	\affil[2]{\textit{\footnotesize Departamento de Física, Universidade Federal Rural do Rio de Janeiro, BR 465-07, CEP 23890-971, Seropédica, RJ, Brazil.}}
	\affil[ ]{}
	\affil[ ]{\footnotesize e-mail: \texttt{olavoabla1@gmail.com, mariojr@ufrrj.br}}

	\maketitle
\begin{abstract}

Poisson electrodynamics is the semi-classical limit of $U(1)$ non-commutative gauge theory. It has been studied so far as a theoretical model, where an external field would be the source of the non-commutative effects in space-time. Being the Standard Model of fundamental interactions a local theory, the prediction of observables within it would be drastically altered by such effects. The natural question that arises is: how do particles interact with this field ? In this work, we will answer this question using point-like charged particles interacting with the Poisson gauge field, investigating how their trajectories are affected using the $\kappa$-Minkowski structure. The interaction arises from the construction of a gauge-invariant action. Using the field solutions, we find the second-order equation for the deformed Lorentz force, indicating possible effects of an emergent gravity due to non-commutativity.    

\end{abstract}
\section{Introduction}\label{1}

The concept of non-commutativity dates back to the early days of quantum mechanics. Bronstein \cite{Bronstein} investigated the quantum effects of weak gravitational fields by asking how small the uncertainty $\Delta x$ in a particle's position can be, given that general relativity states that for each mass or energy, the curvature of space-time is created, which is proportional to the Schwarzschild radius, $r_S \sim (E/M_{P}) \, \ell_{P}$ \cite{Nekrasov}. This suggests that space-time becomes non-local at the Planck scale, {\it i.e.}, $\Delta x \, \Delta r_S \geq \ell_P^2$ \cite{DoplicherPLB,DoplicherCMP}, and one may impose by hands that the quantum gravity effects are given by an external field, in which the observables of the space-time are governed by the commutation relation
\begin{equation}
	[\, x^{\mu}\, , \, x^{\nu}\,]=i\,\Theta^{\mu\nu}(x) \; ,
\end{equation}
where $x^{\mu}=(x^{0},x^{i})$ are position operators.
Surprisingly enough, such effects were theoretically obtained in the context of string theory \cite{seibergwitten99}--\cite{brane2}, quantum electrodynamics \cite{ChaichianQM}--\cite{NCQED}, $3$-dimensional field theory \cite{Vitale2013,Vitale2014}, and effective field theory of quantum gravity \cite{su(2)star3} using the star product approach \cite{Kontsevich,KV2008}. Although the non-commutative formalism had been extensively investigated over the last twenty years \cite{SzaboReview}, some aspects remain not fully understood, specially when one considers position-dependent non-commutativity parameters, particularly in gauge theories \cite{PhysRep}--\cite{rho2}. Mathematical tools have been developed in an attempt to address some of these unsolved issues, such as the use of higher algebraic structures \cite{BBKL}--\cite{kup24-2} and the symplectic realizations
\cite{Kupriyanov:2018yaj}--\cite{BKK}. 

These studies, using the semi-classical approach, developed a new way to introduce non-commutative gauge theories, known as {\it Poisson electrodynamics} \cite{Kupriyanov:2021aet}. On the last year, the investigation using symplectic groupoids \cite{Kup24,Cosmo} brought the possibility of introduce interactions in the theory, through point-like particle examples on $\mathfrak{su}(2)$ \cite{Kup2024} and $\lambda$-Minkowski \cite{BKK} structures. The aim of this work is to make progress in this direction, using the formulation on $\kappa$-Minkowski space-time. We construct the equations of the electrodynamics in $\kappa$-Minkowski space-time, and obtain the solution for the potential and electric field in the case of static charges. Thereby, we study the problem of a charged particle interacting with the central force associated with the $\kappa$-electrostatic field.

This paper is outlined as follows: In Sections $2$, the action principle with gauge invariance is presented, which allows the construction of the dynamics of a particle interacting with a Poisson gauge field \cite{Kup24}. In Section $3$, the Poisson electrodynamics and the field equations are derived in the context of $\kappa$-Minkowski space-time \cite{KKV3}. The solutions for the potential and electric field of a point-like charge are also obtained. In Section $4$, we derive the equations of motion for a charged particle interacting with the Poisson gauge field. Finally, in Section $5$, we study the trajectories under a central force \cite{Kup2024}.
We use the natural units of $\hbar=c=1$ throughout this paper. The greek index $\mu,\nu=\{ \, 0,1,2,3 \, \}$ correspond to the usual space-time signature, with the metric tensor given by $\eta^{\mu\nu}=(+1,-1,-1,-1)$ used to raise and low these indexes. The index $i,j,k=\{1,2,3\}$ are used to spatial coordinates.

\section{Action principle and minimal coupling}

Our goal in this section is to review the construction of the gauge invariant action, for a particle minimally coupled to the Poisson gauge field, following the work \cite{BKK}. From this, we will study the resulting equations of motion. First, let's recall that, in the case of commutative electrodynamics, the motion of a charged relativistic scalar particle, minimally coupled to an external electromagnetic field $A_{\mu}$, is described by the first-order action
\begin{equation}
	S_0=-\int d\tau\left[ \, p_{\mu}\dot{x}^{\mu}+\lambda\,(P^{\mu}P_{\mu}-m^2) \, \right] \; ,
\end{equation}
where $\lambda$ is a Lagrange multiplier, $\tau$ is the proper time, and $P_{\mu} = p_{\mu} - A_{\mu}$ is the modified linear momentum by the minimal coupling, 
or the gauge invariant momentum. It is easy to see that the set of gauge transformations
\begin{equation}
	\delta^0_gA_{\mu}=\partial_{\mu}g \; , \quad \delta^0_gx^{\mu}=\{g,x^{\mu}\}_0=0\quad\text{and}\quad
	\delta_g^0p_{\mu}=\{g,p_{\mu}\}_0=0 \, ,
\end{equation}
for the gauge parameter $g$, make the first-order action invariant, $\delta^0_g S_0 = 0$, with $[\delta^0_g, \delta^0_h] = 0$, where\footnote{for any matrix the upper index enumerates rows, whilst the lower one enumerates columns.} $\{ x^{\mu}, x^{\nu} \}_0 = \{ p_{\mu}, p_{\nu} \}_0 = 0$ and $\{ x^{\mu}, p_{\nu} \}_0 = \delta^{\mu}_{\nu}$ are the canonical Poisson brackets \cite{Woodhouse}. To build the first-order action of Poisson electrodynamics, we look for a particle model described by an action $S[x,p]$, which must satisfy two conditions:
\begin{itemize} 
	\item This action must, from its variation and the consequent Euler-Lagrange equations, generate the Poisson brackets
	\begin{equation} \label{PB1}
		\{ x^{\mu}, x^{\nu} \} = \Theta^{\mu\nu}(x) \; , \quad \{ x^{\mu}, p_{\nu} \} = \gamma^{\mu}_{\nu}(x,p)\,,\quad\{ p_{\mu}, p_{\nu} \} = 0 \; , 
	\end{equation}
	so that $S[x,p] = S_0[x,p] + \mathcal{O}(\Theta)$.
	\item This action must be invariant under the Poisson gauge transformations
	\begin{equation} 
		\delta_g A_{\nu} = \gamma^{\mu}_{\nu}(A)\,\partial_{\mu} g(x) + \{ \, A_{\nu}(x) \, , \, g(x) \, \} \; , 
	\end{equation}
	{\it i.e.}, $\delta_g S[x,p] = 0$. 
\end{itemize}
We will now describe the necessary mathematical tools to present the interaction of a charged point particle with the Poisson gauge field. For simplicity, we first assume that the non-commutativity is of the Lie algebra type, with its components given by $\Theta^{\mu\nu}(x) = f^{\mu\nu}_{\rho} \, x^{\rho}$, and the objects $f^{\mu\nu}_{\rho}$ are the structure constants of a Lie algebra $\mathfrak{g}$. In this case, the symplectic realization, $\{x^{\mu},x^{\nu}\}=2\,\alpha\,\varepsilon^{\mu\nu}_{\rho}\, x^{\rho}$, being $\alpha$ a constant real number, and the Jacobi identity on the Poisson brackets (\ref{PB1}) implies that the $\gamma^{\mu}_{\nu}$-matrices must satisfy the equation \cite{Kupriyanov:2018yaj}
\begin{equation}\label{eq2}
	\gamma^{\mu}_{\rho}\,\partial^{\rho}_p\gamma_{\nu}^{\sigma}-\gamma^{\sigma}_{\rho}\,\partial^{\rho}_p\gamma^{\mu}_{\nu}-\gamma^{\rho}_{\nu}\,f^{\mu\sigma}_{\rho}=0 \; .
\end{equation}
By introducing the inverse matrix, $\bar{\gamma}^{\mu}_{\nu} := (\gamma^{-1})^{\mu}_{\nu}$, we can verify that (\ref{eq2}) implies into the relation
\begin{equation}\label{gammabar}	\partial^{\mu}_p\,\bar\gamma^{\sigma}_{\nu}-\partial^{\sigma}_p\,\bar\gamma^{\mu}_{\nu}+\bar\gamma^{\mu}_{\rho}\,f^{\rho\beta}_{\nu}\,\bar\gamma^{\sigma}_{\beta}=0 \; ,
\end{equation}
which is a Maurer-Cartan equation $d \bar\gamma_{\mu} = -\frac{1}{2} \, f_{\mu}^{\nu\rho} \, \bar\gamma_{\nu} \wedge \bar\gamma_{\rho}$ for the 1-form, and ${\bar\gamma} = \bar\gamma_{\mu} \, dp^{\mu}$ \cite{Kup24}. The notation $\partial^{\mu}_p$ means the derivatives in relation to momentum components. Now, suppose the generalized coordinates $(q^{\mu}, p_{\nu})$ are defined such that the Poisson brackets are canonical, $\{ q^{\mu}, q^{\nu} \}=\{ p_{\mu}, p_{\nu} \} = 0$, and $\{ q^{\mu} \, , \, p_{\nu} \} = \delta^{\mu}_{\nu}$, 
which means these are Darboux coordinates \cite{Woodhouse}. Thereby, the representation of the original coordinates in terms of Darboux one is given by, $x^{\mu}=q^{\nu}\,\gamma^{\mu}_{\nu}$.

One may observe that the integration measure for structures of the type $\mathfrak{su}(2)$ is trivially equal to one, which guarantees an invariant free particle action \cite{Kup24}
\begin{equation}\label{action}
	S=\int\! d\tau \,p_{\mu}\,\dot q^{\mu}=-\int\! d\tau\, \dot p_{\mu} \, \bar\gamma^{\mu}_{\nu}\, x^{\nu} \, .
\end{equation}
Introducing the interaction, we obtain the action
\begin{equation}\label{4a}
	S=-\int d\tau \, \left[ \, \dot p_{\mu}\,\bar\gamma^{\mu}_{\nu}\, x^{\nu}+\lambda \, H(x,p) \, \right] \; ,
\end{equation}
The variation of the action leads to the equations of motion
\begin{equation}\label{eom}
	\dot x^{\mu}=\lambda \, \{x^{\mu},H\}
	\, , \qquad \mbox{and} \qquad 
	\dot p^{\mu}=\lambda \, \{p^{\mu},H\} \, .
\end{equation}
One can verify the conditions for gauge invariance by first observing how equation (\ref{4a}) transforms under gauge transformations \cite{BKK}
\begin{subequations}
	\begin{eqnarray}\label{5}
		\delta_g x^{\mu} &=& \{ \, g(x) \, , \, x^{\mu} \, \}=-f^{\mu\nu}_{\rho}\,x^{\rho}\,\partial_{\nu}g(x) \; ,
		\\ 
		\delta_g p_{\mu} &=&\!\! \{ \, g(x) \, , \, p_{\mu} \, \}=\gamma_{\mu}^{\nu}(p)\,\partial_{\nu}g(x) \; .
	\end{eqnarray}
\end{subequations}
Using such conditions, we can rewrite the action with the interaction Hamiltonian (\ref{4a}) as follows

\begin{equation}\label{5a}			
\delta_gS_0=-\int\!d\tau\big{[}\frac{d}{d\tau}(\partial_{\mu}g)\,x^{\mu}+\dot{p}_{\mu}\left(\bar\gamma^{\sigma}_{\beta}
	\,\partial^{\mu}_p\gamma^{\beta}_{\xi}\,\bar\gamma^{\xi}_{\nu}+\partial^{\sigma}_p\bar\gamma^{\mu}_{\nu}-\bar\gamma^{\mu}_{\beta}\,f^{\beta\xi}_{\nu}\,\bar\gamma^{\sigma}_{\xi} \right)\gamma^{\rho}_{\sigma}\,\partial_{\rho}g\,x^{\nu}+\delta_gH\big{]} \, .
\end{equation}
Using equation (\ref{gammabar}) in the second term and integrating, we represent the right-hand side as
\begin{equation}
	\int dg-\int d\tau \, \delta_gH \; .
\end{equation}
This actually means that the transformations (\ref{5}) do not affect the dynamics if the corresponding Hamiltonian remains invariant under such gauge transformations, $\delta_g H = 0$. Note that we do not need the precise expression of the symplectic potential for our current purposes, although it can be constructed \cite{Kupriyanov:2018yaj}. This means that, for the interaction case (\ref{4a}), one can promote the coordinates of the canonical phase space to coordinates invariant under gauge transformations, following \cite{BKK}
\begin{subequations}
	\begin{eqnarray}\label{ic}
		\hspace{-0.5cm}&& x^{\mu} \, \longmapsto \, y^{\mu}(x,A)=x^{\mu}+{\cal O}(\Theta) \; ,
		\\
		\hspace{-0.5cm}&& p_{\mu} \, \longmapsto \, \pi_{\mu}(x,p,A)=p_{\mu}-A_{\mu}(x)+{\cal O}(\Theta) \; ,
	\end{eqnarray}
\end{subequations}
which implies
\begin{subequations}
	\begin{eqnarray}
		\delta_g x^{\mu} \!&=&\! \{g(x),x^{\mu}\}=-\Theta^{\mu\nu}(x)\,\partial_{\nu}g(x) \; ,
		\label{pgt}
		\\ 
		\delta_g p_{\mu} \!&=&\! \{g(x),p_{\mu}\}=\gamma_{\mu}^{\nu}(x,p) \,\partial_{\nu}g(x) \; , \\
		\delta_g A_{\mu}(y) \!&=&\! \gamma_{\mu}^{\nu}(y,A)\,\partial_{\nu}g(y)+\{A_{\mu}(y),g(y)\} \; .
		\label{pgtc}
	\end{eqnarray}
\end{subequations}
It can be noted that \cite{Kup24}
\begin{equation}\label{gtAx}
	\delta_g A_{\mu}(x)=\left.\delta_g A_{\mu}(y)\right|_{y=x}+\partial_{\nu}\,A_{\mu}(x)\,\delta_gx^{\nu}=\gamma_{\mu}^{\nu}(x,A)\,\partial_{\nu}g(x) \; .
\end{equation}
The construction of a invariant Hamiltonian begins with the modification of the canonical covariant momenta, 
with corrections in the non-commutative $\Theta$-parameter, so that it must be invariant under the Poisson gauge transformations (\ref{pgt})-(\ref{pgtc})
\begin{equation}\label{gic}
	\delta_g\,\pi_{\mu}=0 \; .
\end{equation}
The condition (\ref{gic}) yields the following equation
\begin{equation}\label{mepi}
\gamma^{\nu}_{\sigma}(x,p)\,\partial^{\sigma}_p\pi_{\mu}(x,p,A)+\gamma^{\nu}_{\sigma}(x,A)\,\partial^{\sigma}_A\pi_{\mu}(x,p,A)+\Theta^{\nu\sigma}(x)\,\partial_{\sigma}\,\pi_{\mu}(x,p,A)=0 \; .
\end{equation}
Taking the expression for the symplectic realization \cite{Kupriyanov:2018yaj}
\begin{equation} 
\gamma^{\nu}_{\mu}(x,p) = \sum_{n=0}^\infty \gamma_{\mu }^{\nu (n)}=\delta^{\nu}_{\mu}-\frac{1}{2} \partial_{\mu} \Theta^{\nu\sigma} p_{\sigma}-\frac{1}{12}\left(2\Theta^{\sigma\rho}\partial_{\mu}\partial_{\rho}\Theta^{\beta\nu}
	+\partial_{\mu}\Theta^{\beta\rho}\partial_{\rho}\Theta^{\nu\sigma}\right)p_{\sigma}p_{\beta}+{\cal O}(\Theta^3),
\end{equation}
we obtain
\begin{eqnarray}
	\pi_{\mu}(x,p,A)&=& p_{\mu}-A_{\mu}+\frac{1}{2}\, \partial_{\mu} \Theta^{\nu\sigma}\, A_{\nu}\, p_{\sigma}\notag\\
	&&+\frac{1}{12}\left(2\,\Theta^{\sigma\rho}\partial_{\mu}\partial_\rho\Theta^{\nu\beta}+\partial_{\mu}\Theta^{\nu\rho}\partial_{\rho}\Theta^{\beta\sigma}\right)\left(p_{\nu}p_{\sigma}A_{\beta}-A_{\nu}A_{\sigma}p_{\beta}\right)+{\cal O}(\Theta^3) \, .
\end{eqnarray}
By induction, the existence of the solution $\pi_{\mu}^{(n)}(x,p,A)$ to equation (\ref{mepi}) is given for each order $n$, and follows from the existence of the solution $\pi_{\mu}^{(m)}(x,p,A)$ in the previous orders, $m < n$, and the existence of the symplectic realization $\gamma^{\nu}_{\mu}(x,p)$, which was demonstrated in \cite{Kupriyanov:2018yaj}. Some simplifications occur if we work with linear structures. In this case, both the symplectic realization $\gamma^{\nu}_{\mu}(p)$ and the covariant momentum $\pi_{\mu}(p,A)$ do not explicitly depends on the coordinates, with equations (\ref{ic}) and (\ref{mepi}) written as \cite{BKK}
\begin{equation}\label{mepi1}
	\gamma^{\nu}_{\sigma}(p)\,\partial^{\sigma}_p\pi_{\mu}(p,A)+\gamma^{\nu}_{\sigma}(A)\,\partial^{\sigma}_A\pi_{\mu}(p,A)=0 
	\, ,\quad\pi_{\mu}(p,A)=p_{\mu}-A_{\mu}+{\cal O}(\Theta) \, .
\end{equation}
Therefore, the first-order correction of $\pi_{\mu}$ is
\begin{equation}\label{pi}	
	\pi_{\mu}^{(1)}(p,A)=\frac{1}{2}\left[\,\gamma^{{\nu}(1)}_{\mu}(A)\,p_{\nu}-\gamma^{{\nu}(1)}_{\mu}(p)\,A_{\nu}\,\right] \; .
\end{equation}
We will see how this mathematical approach applies in the context of $\kappa$-Minkowski electrodynamics \cite{KKV3}.

\section{The $\kappa$-Minkowski electrodynamics}

Now, we will review the $\kappa$-Minkowski \cite{kappa}--\cite{kappa8} electrodynamics, using some of the results already obtained in works \cite{KKV3}. 
The commutation relations for the position and momentum components are given by
\begin{equation}
	\{x^0,x^i\}=\kappa\,x^i, \quad \; 
	\{x^0,p_i\}=-\kappa\,p_i \,,\quad
	\{x^0,p_0\}=1 \; , \quad\quad	\{x^i,p_j\}=\delta^i_{j} \, ,
\end{equation}
where the structure constants are defined as
\begin{equation}\label{fkappa}
	f^{\mu\nu}_\sigma = \kappa\,\big(\delta_0^\mu\, \delta^\nu_\sigma - \delta_0^\nu\,\delta^\mu_\sigma \big) \; .
\end{equation} 
We start with the definition of Poisson gauge transformations $\delta_g A_{\mu}$, where $A^{\mu} = (A_0, {\bf A})$ is the gauge field, 
so that the closure of the algebra
\begin{equation}\label{ga} 
	[\,\delta_g\,,\,\delta_h\,]\,A_{\mu} = \delta_{\{g,h\}} \, A_{\mu} \; ,	
\end{equation}
and the correct commutative limit 
\begin{equation}\label{commga}
	\lim_{\Theta\to0}\delta_gA_{\mu} = \partial_{\mu} g \; ,
\end{equation}
are satisfied. Consistency with the semi-classical limit imposes that the gauge covariant derivative acting on a $\varphi$-field \cite{KKV2}
\begin{equation}\label{w6}
	{\cal D}_{\mu}\varphi=\rho_{\mu}^{\nu}(A)\,\gamma_{\nu}^{\sigma}\left(A\right)\partial_{\sigma}\varphi+\rho_{\mu}^{\nu}\left(A\right)\left\{\,A_{\nu}\,,\,\varphi\,\right\} \; ,
\end{equation}
and the matrices $\gamma(A)$ and $\rho(A)$ are given by
\begin{equation}\label{gamma-kappa}
	\gamma(A^{i})  =  \left(
	\begin{array}{cccc}
		1 &-\kappa A_1 &-\kappa A_2 & -\kappa A_3 \\
		0 &1 &0 &0 \\
		0 &0 &1 &0 \\
		0 &0 &0 &1 
	\end{array}
	\right) \; .
\end{equation}
and
\begin{equation}\label{rho}
	\rho(A_{0})  =  \left(
	\begin{array}{cccc}
		1 &0 &0 & 0 \\
		0 & \e^{\kappa A_0} &0 &0 \\
		0 &0 &\e^{\kappa A_0} &0 \\
		0 &0 &0 &\e^{\kappa A_0} 
	\end{array}
	\right) \; ,
\end{equation}
respectively. The Poisson strength tensor is \cite{KKV2}
\begin{equation}\label{pfs}
	{\cal F}_{\mu\nu}=\rho^{\alpha}_{\mu}\rho^{\beta}_{\nu}\left(\gamma_{\alpha}^{\sigma}\,\partial_{\sigma} A_{\beta}-\gamma_{\beta}^{\sigma}\,\partial_{\sigma} A_{\alpha}+\{ \, A_{\alpha} \, , \, A_{\beta} \, \}\right) \; .
\end{equation}

It should be stressed that the Poisson strength tensor given by (\ref{pfs}) is not the only possible form in Poisson electrodynamics. Other strength tensors can be constructed within the symplectic groupoid approach, as discussed in \cite{Kup24}: the fully covariant strength tensor $F^s$ and the gauge-invariant one $F^t$. For constant non-commutativity, the precise relationship was constructed in \cite{Kup24}. For other structures, such as the linear one (in particular for the $\kappa$-Minkowski case), the relation between the fully covariant and gauge invariant strength tensors remains unknown. Understanding this relation would be valuable, and we plan to investigate it, following the prescription outlined in \cite{Cosmo}.

We consider the equations of motion given by the first pair of the Maxwell-Poisson equations, obtained by the non-Lagrangian formalism \cite{KKV3}
\begin{equation}\label{source}
	{\cal D}_{\mu}{\cal F}^{\mu\nu}+\frac{1}{2}\,{\cal F}_{\sigma\beta}f^{\sigma\beta}_{\mu}\,{\cal F}^{\nu\mu}-{\cal F}_{\sigma\beta}f^{\nu\beta}_{\mu}\,{\cal F}^{\sigma\mu}+\,\alpha\left(\,f^{\mu\nu}_{\mu}\,{\cal F}^2+4f^{\mu\sigma}_{\mu}\,{\cal F}_{\sigma\beta}{\cal F}^{\beta\nu}\,\right)=J^{\nu} \; ,
\end{equation}
in which $\alpha$ is a constant real parameter, $J^{\mu} = (\rho, {\bf J})$ is a classical current, and $f^{\mu\nu}_{\sigma}$ is the Lie algebra structure constant (\ref{fkappa}). Using the definition (\ref{fkappa}), the source equation (\ref{source}) is reduced to
\begin{equation}
	{\cal D}_{\mu}{\cal F}^{\mu\nu}-\kappa(1+3\alpha)\delta^{\nu}_0{\cal F}^2-2\kappa(1+6\alpha){\cal F}_{0\mu}{\cal F}^{\mu\nu}=J^{\nu}\,.
\end{equation}

The Bianchi identity leads to the second pair of Maxwell-Poisson equations without sources
\begin{equation}
	{\cal D}_{\mu}{\cal F}_{\nu\sigma}-{\cal F}_{\mu\beta}\,f^{\beta\zeta}_{\nu}\,{\cal F}_{\zeta\sigma}+\text{cycl.}(\mu\nu\sigma)=0 \; .
\end{equation}
The commutative limit leads to the standard Maxwell's equations. Now, we present a new solution, based on the results obtained in \cite{KKV3}. In the electrostatic case, the potentials are written as $A_0 = V({\bf r})$ and ${\bf A} = {\bf 0}$, and the field equations reduce to the Poisson equation, namely
\begin{equation}\label{EqV}\, \nabla^2V+\kappa(1+6\alpha) \,(\nabla V)^2=-\rho\,\e^{-2\kappa V} \; .
\end{equation}
For $\alpha=-1/12$, the authors on \cite{KKV3} considered such case to explore the deformed plane-wave solutions. Considering the problem of a point-like charge $(Q)$, with $\rho({\bf r}) = Q\,\delta^3({\bf r})$, the potential has only radial dependence in spherical coordinates, where equation (\ref{EqV}) reduces to
\begin{equation}\label{EqVradial}
	\frac{1}{r^2}\frac{d}{dr}\left(r^2\,\frac{dV}{dr}\right)+\kappa(1+6\alpha) \,\,\left(\frac{dV}{dr}\right)^2=0 \; ,
\end{equation}
for $r \neq 0$. The corresponding solution to (\ref{EqVradial}) is
\begin{equation}\label{PotV}
	V(r)=\frac{1}{\kappa(1+6\alpha) }\ln\left[ 1+(1+6\alpha)\frac{\kappa Q}{r} \right] \; .
\end{equation}
The deformed electrostatic field is
\begin{equation}
	{\bf E}(r)=\frac{Q}{r^2}\left[\frac{r}{r+(1+6\alpha)\kappa Q}\right]\hat{{\bf r}} \; .
\end{equation}
Note that, the limit $\kappa \rightarrow 0$, as well as $\alpha=-1/6$, one recovers the usual Coulomb electrostatic results. For $\kappa \neq 0$, the behavior of the potential and the electrostatic field changes with the choice of the $\alpha$-parameter, providing a family of electrostatics theories.

Let us now study the interaction of the Poisson gauge field with point particles. We use the equation obtained in \cite{BKK} for the $\kappa$-Minkowski structure. It can be verified that the components of the gauge covariant momenta $\pi^{\mu}$, given by equation (\ref{pi}) are
\begin{equation}\label{gimomenta}
	\pi_{\mu}(p,A) =\big[ \, w\, \rho_{\mu}^{\nu}(A) + (1-w)\, \rho_{\mu}^{\nu}(p) \, \big](p_{\nu}-A_{\nu}) \; ,
\end{equation}
where $w$ is a real parameter. Equation (\ref{gimomenta}) satisfies the third equation in \eqref{mepi1}, with $\rho_{\mu}^{\nu}$ given by (\ref{rho}). The covariant momenta introduce the minimal coupling to the Hamiltonian
\begin{equation}
	H=H_0(\pi^2)+A_0(x) \; ,
\end{equation}
in which $A_0(x)$ is the scalar potential that we have just obtained as a specific solution in (\ref{PotV}). By construction, $H_0(\pi^2)$ is invariant under gauge transformation. The action of the relativistic particle is
\begin{equation}\label{action-kappa}
	S=-\int d\tau \left[ \, x^{\mu}\,\bar\gamma_{\mu}^{\;\,\nu}(p)\,\dot p_{\nu}+\lambda\left(\pi_{\mu}\,\pi^{\mu}-m^2\right) \, \right] \; ,
\end{equation}
with all the necessary objects defined earlier, and the components of the covariant gauge momentum $\pi^{\mu} = (\pi_{0}, \boldsymbol{\pi})$ are given according to (\ref{gimomenta}), with $w = 1$, that is
\begin{equation}
	\pi_0=p_0-A_0(x) \; \; , \;\; \pi_i=\e^{\kappa A_0(x)}\left[p_i-A_i(x)\right] \; .
\end{equation}
In this case, the derivatives with respect to the components $A^{\mu} = (A^{0}, A^{i})$ are given by
\begin{equation}
	\partial_A^0\pi_0=-1\,,\quad\partial_A^i\pi_0=0\,,\quad
	\partial_A^0\pi_i=\kappa\,\pi_i\,,\quad \partial^i_A\pi^j=-\e^{\kappa A_0(x)}\,\delta^{ij}\,,
\end{equation}
and the following relations can be obtained
\begin{subequations}
	\begin{eqnarray}
		(\partial_A^0\pi_\mu)\,\pi^\mu &=& -\pi^0-\kappa \, {\bf \pi}^2 \; ,
		\\
		(\partial_A^i\pi_\mu)\,\pi^\mu &=& \e^{\kappa A_0(x)}\,\pi^i \; .
	\end{eqnarray}
\end{subequations}
Considering that the potential $A^{\mu}$ does not depend on the time coordinate (static case), 
the first-order equations resulting from (\ref{action-kappa}) are written below
\begin{subequations}
	\begin{eqnarray}
		\dot x^0&=& 2\,\lambda\big{[}(1-\kappa\,x^j\partial_jA_0)(\pi^0+\kappa \, {\bf \pi}^2)
		+\kappa\,\e^{\kappa A_0}\,(A_j+x^m\partial_mA_j)\,\pi^j\big{]} \, ,
		\\
		\dot x^i&=& -2\,\lambda\,\e^{\kappa A_0}\pi^i \, ,
		\\
		\dot \pi_0&=& -2\,\lambda\,\e^{\kappa A_0}\,(\partial_iA_0)\,\pi^i \, ,
		\\
		\dot\pi_i &=& 2\,\lambda\,\e^{\kappa A_0}\,(\partial_iA_0)\,(\pi^0+\kappa \, {\bf \pi}^2)+2\,\lambda\,\e^{2\kappa A_0}\,F_{ij}\,\pi^j
		-2\,\lambda\,\kappa\,\pi_i\,\e^{\kappa A_0}\,(\partial_jA_0)\,\pi^j \, ,
	\end{eqnarray}
\end{subequations}
with the condition of $\pi_{\mu} \, \pi^{\mu} = m^2$. The first integral of motion is obtained directly, as we know that the Hamiltonian does not depend on the time coordinate for the static case, which is $p_0$. It can be interpreted as the energy of the system, considering the gauge fixing of $\lambda = 1/2$, we obtain
\begin{equation}
	p_0=\pm \, \sqrt{\e^{2\kappa A_0}(\bold{p}-\bold{A})^2+m^2}+A_0 \; .
\end{equation}
The deformed Lorentz force is given by
\begin{equation}\label{soeom-kappa}
	\ddot x^i=2\,\kappa\,\partial_jA_0\,\dot x^j\,\dot x^i-(\partial_iA_0)\,\frac{\dot x^0\,\e^{2\kappa A_0}}{1-\kappa\,x^j\partial_jA_0}-\e^{2\kappa A_0}\left[F_{ij}-\kappa\,(\partial_iA_0)\,\frac{A_j+x^m\partial_mA_j}{1-\kappa\,x^m\partial_mA_0}\right]\dot x^j \; .
\end{equation}
In the particular case in which the scalar potential is zero, these equations reduce to the standard Lorentz force, {\it i.e.}, $\ddot{x}^i = -F^{ij} \, \dot{x}^j$. 
One may rewrite this expression as
\begin{equation}
	\ddot{x}^i+\Gamma^i_{jl}\,\dot x^j\,\dot x^l=f^i \; ,
\end{equation}
which suggests an emergent gravity \cite{Szabo}--\cite{szabo2006}. It would be worth investigating this curious effect in future work, particularly following \cite{kappageodesic}, where the authors find a similar {\it gravity like term}. Besides, it would be interesting to understand how it affects constant non-commutativity \cite{Kup24}, since this effect has not been observed in other types of non-commutative electrodynamics \cite{NCED3}--\cite{Liang}.

Now, we will treat the dynamics of a charged particle interacting with the Poisson gauge field. Considering ${\bf A} = {\bf 0}$, $A_0 = V({\bf r})$, and the parametrization of the worldline, $x^0 = c_\kappa \, \tau$, we obtain the second-order equation
\begin{equation}\label{eqmotion}
	\frac{d}{d\tau}[ \, \e^{-2\kappa V({\bf r})} \,  {\bf v}\,]=\frac{-\,c_\kappa \, \nabla V}{1-\kappa \, {\bf r} \cdot \nabla V} \; ,
\end{equation}
with ${\bold v}$ being the velocity of the particle, and $(\nabla V)$ the gradient of the potential obtained in the eq. (\ref{PotV}).
\section{The orbit equation}
The components of the angular momentum
\begin{equation}
	L_i=\varepsilon_{ijk} \,q_j \, p_k \; ,
\end{equation}
can be written in terms of the Darboux coordinates, $q^{\mu} = \bar{\gamma}^{\mu}_{\nu} \, x^{\nu}$. For the spatial components, $q^i = x^i$, where $\bar{\gamma}^i_{j} = \delta^i_{j}$. This leads to the conservation law
\begin{equation}
	\frac{d}{d\tau}[ \, \e^{-2\kappa \, V({\bf r})}  \, {\bf r}  \times {\bf v} \, ]=0 \; ,
\end{equation}
whose conserved quantity is defined as the deformed angular momentum
\begin{equation}\label{Lconserv}
	{\bf L}= \e^{-2\kappa \, V({\bf r})}  \, {\bf r}  \times {\bf v} \; .
\end{equation}
For any central potential, the angular momentum satisfies the plane equation $ {\bf r} \cdot {\bf L} = 0 $. The cross product of (\ref{eqmotion}) with the angular momentum (\ref{Lconserv}) leads to the following result
\begin{equation}\label{relrunge}
	\frac{d}{d\tau} \left[ \, \e^{-2\kappa \, V({\bf r})} \, {\bf v} \times {\bf L}  \, \right]=\e^{-2\kappa \, V({\bf r})} \, f(r) \, r^2 \frac{d}{d\tau}\left( \frac{{\bf r}}{r} \right) \; ,
\end{equation}
in which the central force $f(r)$ from (\ref{eqmotion}) associated with the potential (\ref{PotV}) is given by
\begin{equation}\label{fr}
	f(r)= \frac{c_{\kappa}Q}{r^2}\left[1+(1+3\alpha)\,\frac{2\kappa Q}{r} \, \right]^{-1} \; .
\end{equation}

The relation (\ref{relrunge}) shows that it is not possible to obtain a conserved Laplace-Runge-Lenz vector for the central force (\ref{fr}). It is important to note that the authors in \cite{Kup2024} obtained a super-integrable model for the $\mathfrak{su}(2)$ non-commutative Poisson structure, modifying the Hamiltonian. Thus, the conservation of the Laplace-Runge-Lenz vector is recovered only in the commutative limit, $\kappa \rightarrow 0$.
\begin{figure}
	\centering
	\includegraphics[width=0.75\textwidth]{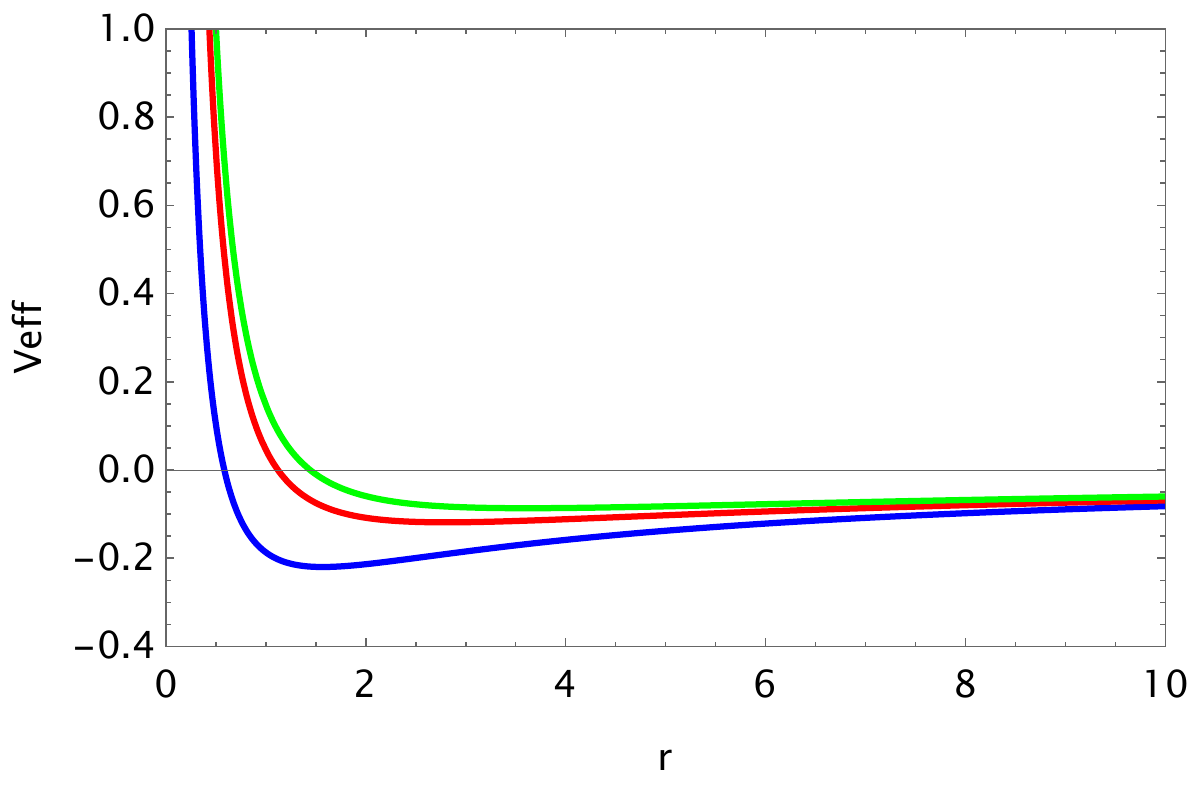}
	\caption{\label{fig:diagram}
		The effective potential as a function of radial distance. We plot this figure for $L = 1$, $Q = -1$, $\kappa = 0.5$ (in length units), and $c_{\kappa} = 0.99$, when $\alpha = -1$ (blue line), $\alpha = -2$ (red line), and $\alpha = -3$ (green line). The effective potential has two return points for a negative energy in the curve of $\alpha = -1/2$.}
\end{figure}
Since the potential (\ref{PotV}) and the force (\ref{fr}) have radial symmetry, we explicitly write the equations of motion (\ref{eqmotion}) in spherical coordinates $(r, \theta, \varphi)$
\begin{subequations}
	\begin{eqnarray}
		&&\ddot{r}-r\,\dot{\theta}^2-r\sin^2\theta\,\dot{\varphi}^2-2\kappa\dot{r}^2\,\partial_rV=
		f(r) \, \e^{2\kappa V(r)} ,
		\label{coordreq}
		\\
		&&\frac{d}{dt}\left[\,\e^{-2\kappa\,V(r)}\,r^2\,\dot{\theta}\,\right]=r^2\sin\theta\cos\theta\,\dot{\varphi}^2\,\e^{-2\kappa\,V(r)} ,
		\label{coordthetaeq}
		\\
		&&\frac{d}{dt} \left[ \, \e^{-2\kappa\,V(r)}\,r^2\, \sin^2\theta\,\dot{\phi} \, \right]=0.
		\label{coordphieq}
	\end{eqnarray}
\end{subequations}
The equation (\ref{coordphieq}) implies that the canonical momentum associated with the coordinate $\phi$ is conserved
\begin{figure}
	\centering
	\includegraphics[width=0.75\textwidth]{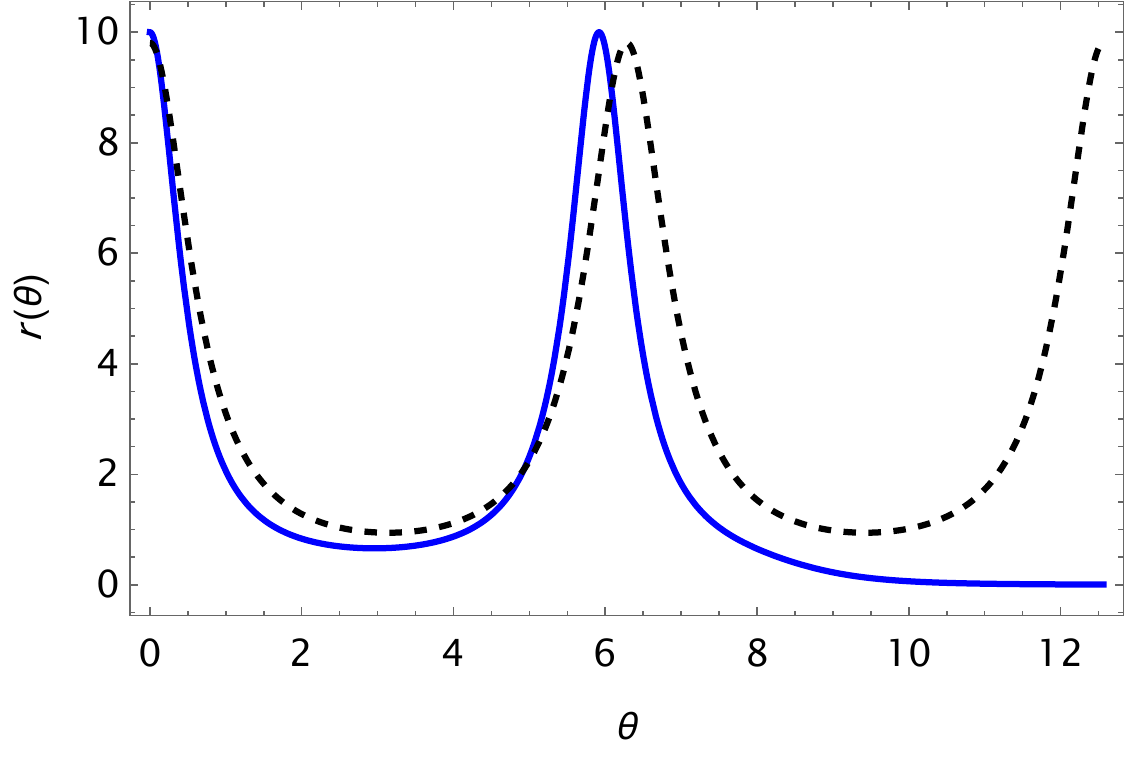}
	\caption{\label{fig3} The numerical solution for the orbital equation with $\kappa \neq 0$ is represented by the blue line. The dashed black line represents the analytical solution of the usual orbital equation, when $\kappa \rightarrow 0$. We chose the parameters as $L=1$, $Q=-1$, $\alpha=-1$, $\kappa=0.5$, and $c_{\kappa}=0.99$ to plot the curves.}
\end{figure}

\begin{equation}
	p_{\phi}=\e^{-2\kappa\,V(r)}\,r^2\, \sin^2\theta\,\dot{\phi}=\mbox{constant} \; ,
\end{equation}
because it is a cyclic coordinate \cite{Arnold}. If we consider the particle's motion in a particular plane, such that $\theta = 0$, $p_{\phi} = 0$, consequently, $\phi$ is constant. Thus, the particle moves in a plane where we can choose polar coordinates to describe the motion, as in classical mechanics. In this way, the equations of motion in polar coordinates $(r, \theta)$ lead to the relation
\begin{equation}
	\dot{\theta}=\frac{L}{r^2} \, \e^{ 2 \kappa \, V(r) } \; ,
\end{equation}
being $L$ the magnitude of the conserved angular momentum. Eliminating $\dot{\theta}$, the radial equation is given by,
\begin{equation}\label{radialeq}
	\ddot{r}-2\kappa \, \dot{r}^2 \, \frac{dV}{dr}=F_{eff}(r) \; , 
\end{equation}
where $F_{eff}$ is the effective force
\begin{equation}
	F_{eff}(r) =\frac{L^2}{r^3} \, \e^{4\kappa \, V(r)}+\frac{c_{\kappa} \, Q \, \e^{2\kappa\, V(r)} }{r[\,r+2\kappa Q (1+3\alpha) \,]}  \; .
\end{equation}

The corresponding effective potential $V_{eff}$ is plotted as a function of radial distance in the Fig. $2$. We choose parameters such as $L = 1$, $Q = -1$ (attractive potential), $\kappa = 0.5$ (in length units), and $c_{\kappa} = 0.99$, for $\alpha = -1$ (blue line), $\alpha = -2$ (red line), $\alpha = -3$ (green line). The case for $\alpha = -1$ shows two return points for a given negative energy. The orbital equation is obtained from (\ref{radialeq}), eliminating the parameter $\tau$ in terms of the coordinate $\theta$, and using the transformation $r(\theta) = u(\theta)^{-1}$, the equation for $u(\theta)$ is written as
\begin{equation}\label{eqorbitau}
	\frac{d^2u}{d\theta^2}+u-2\kappa \, \frac{dV}{du} \left( \frac{du}{d\theta} \right)^2 = -\frac{(c_{\kappa} Q/L^2)\, \e^{-2\kappa\,V(u^{-1})}}{1+2\kappa \, Q (1+3\alpha)u(\theta)} \; .
\end{equation}
The limit $\kappa \rightarrow 0$ reduces (\ref{eqorbitau}) to the orbital equation of the Kepler problem, for a central force like $f(r) = Q/r^2$. The numerical solution of (\ref{eqorbitau}) is illustrated in the figure (\ref{fig3}) by the blue line. The dashed black line represents the commutative case. These curves were plotted for $L = 1$, $Q = -1$, $\kappa = 0.5$, and $c_{\kappa} = 0.99$, when the $\alpha$-parameter is $\alpha = -1$ (in the case of $\kappa \neq 0$). Comparing these two solutions shows the behavior of the curves when $\theta > 3\pi$, where the dashed line maintains the periodic motion of the particle.
The blue line decays for $\theta > 3\pi$, where the radial distance goes to zero rapidly. Therefore, the blue curve shows that the orbit is not closed, since the noncommutative $\kappa$-Minkowski space-time does not allow for a conserved Laplace-Runge-Lenz vector \cite{Arnold}, as would be expected to satisfy Bertrand’s theorem. It is noteworthy that this case was not observed when a constant noncommutative parameter was considered \cite{Liang}.

\section{Conclusions}

Using the symplectic groupoid prescription to include interactions \cite{Kup24}, we investigated a family of solutions to Poisson electrodynamics, defined in the $\kappa$-Minkowski space-time. The electrostatic equation for a scalar potential is obtained, along with the corresponding solutions for a point test charge.
We showed that the deformed potentials and the equations obtained in the commutative limit recover the usual objects of Maxwell's electrodynamics. Next, we reviewed the dynamics of a charged particle coupled to the Poisson gauge field \cite{BKK}. The equations of motion in the presence of scalar and vector potentials under the action of the corresponding central force were studied. Since energy and angular momentum are conserved, numerical solutions were obtained for the trajectory equation in a plane, and we compared these solutions with the usual solutions of the commutative case ($ \kappa = 0 $). For $\kappa \neq 0$, the particle's trajectory is not closed \cite{Kup2024}, despite the attractive behavior of the central force, and the effective potential shows return points for a given negative energy. Perspectives for future work include elucidating how the model would be affected by a different choice of Poisson field strength \cite{Kup24,Cosmo}, as well as investigating the new field configurations applied to the model with matter fields \cite{AS24}. Furthermore, it would be worth to explore possible emergent gravitational phenomena that depends on the $\kappa$-parameter. This new term that appeared in (\ref{soeom-kappa}) could describe the motion on a curved trajectory for the particle interacting with the noncommutative field, as appeared in Eq. $(42)$ of \cite{kappageodesic}. Another perspective is to investigate the central force problem for different noncommutative structures, such as $\rho$-Minkowski  \cite{rho}-\cite{FabianoJHEP}. Construct a star product that quantizes such structures may lead to the study of the hydrogen atom \cite{Kup13} and fermionic particles on non-commutative spaces \cite{Kup}.
\section*{Acknowledgements}
We are grateful to Vladislav Kupriyanov for early discussions on this subject and the valuable remarks. 
This project was partially supported by the Coordena\c{c}\~ao de Aperfei\c{c}oamento de Pessoal de N\'ivel Superior (CAPES, Brazil) - Finance Code 001.

\end{document}